\documentclass[prper,longbibliography,reprint,floatfix,superscriptaddress]{revtex4-2}
\usepackage[T1]{fontenc}	
\usepackage[latin9]{inputenc}	
\usepackage{geometry}		
\usepackage{graphicx}
\usepackage[above,below]{placeins}	
\usepackage{times}
\usepackage{xcolor}
\usepackage{textgreek}
\usepackage{amsmath,amssymb}

\usepackage{hyperref}  
\hypersetup{colorlinks=true,urlcolor=blue,citecolor=blue,linkcolor=blue}   
\usepackage{enumitem}          
\setlist{nosep}                 

\begin{document}


\title{Comparing item response theory models for ranking incorrect response options}
\author{Trevor I.\ Smith}
\affiliation{Department of Physics \& Astronomy, Rowan University, 201 Mullica Hill Rd., Glassboro, NJ 08028, USA}
\affiliation{Department of STEAM Education, Rowan University, 201 Mullica Hill Rd., Glassboro, NJ 08028, USA}

\author{Mohammad A.\ Kazmi}
\affiliation{Department of Physics \& Astronomy, Rowan University, 201 Mullica Hill Rd., Glassboro, NJ 08028, USA}

\author{Richard R.\ Sarles, III}
\affiliation{Department of Physics \& Astronomy, Rowan University, 201 Mullica Hill Rd., Glassboro, NJ 08028, USA}

\author{Joshua A.\ Sbrana}
\affiliation{Department of Physics \& Astronomy, Rowan University, 201 Mullica Hill Rd., Glassboro, NJ 08028, USA}

\author{Cody W.\ Soper}
\affiliation{Department of Physics \& Astronomy, Rowan University, 201 Mullica Hill Rd., Glassboro, NJ 08028, USA}

\author{Nasrine Bendjilali}
\affiliation{Department of Mathematics, Rowan University, 201 Mullica Hill Rd., Glassboro, NJ 08028, USA}

\begin{abstract}
    Previous work has shown that item response theory may be used to rank incorrect response options to multiple-choice items on commonly used assessments. This work has shown that, when the correct response to each item is specified, a nominal response model (NRM) may be used to rank the incorrect options. We seek to determine the robustness of these results by applying the NRM to all response choices, without specifying the correct response. We apply these analyses to multiple data sets (each with more than 9,000 response sets), including pre-instruction and post-instruction responses. We find that the rankings generated without specifying the correct response are consistent with the previously published rankings for one data set; however, we find noticeable differences between rankings generated from different data sets. We provide evidence that discrepancies may result from differences in response rates for less commonly chosen responses. 
    
\end{abstract}
\maketitle

\section{Introduction}
Item response theory (IRT) has become an increasingly popular tool for analyzing responses to common research-based assessments. Traditionally, item response theory models the probability that an individual will choose the correct response to a particular multiple-choice item based on a latent trait of the individual and item parameters. The latent trait $\theta$ is sometimes referred to as a person's ``ability'' in a particular area, or otherwise simply the ``person parameter'' of the model. In the context of research-based multiple-choice tests, the latent trait may be interpreted as an individual's overall knowledge or understanding of the topic of the test. 

Wang and Bao used a three-parameter logistic (3PL) model to show that the IRT latent trait strongly correlates with total score for the Force Concept Inventory (FCI \cite{Hestenes1992}) \cite{Wang2010}. Ding and Beichner included IRT analyses in their compendium of quantitative analyses that are useful for understanding student responses to multiple-choice tests \cite{Ding2009}. Yang, Zabriskie, and Stewart used a multidimensional two-parameter logistic (2PL) model to perform a factor analysis of data from student responses to the Force and Motion Conceptual Evaluation (FMCE \cite{Thornton1998}) to identify a substructure of the test \cite{Yang2019}.

The IRT nominal response model (NRM) introduced by Bock has been used to analyze multiple-choice data in ways that incorporate all response options instead of scoring items as simply correct or incorrect \cite{Bock1972, Bock2007, Thissen2010}. Eaton, Johnson, and Willoughby \cite{Eaton2019irt}, and Smith, Louis, Ricci, and Bendjilali \cite{Smith2020rank} used a nested-logit model (2PL-NRM) to rank incorrect responses to items on the FCI and FMCE, respectively. The 2PL-NRM model applies a 2PL model to determine the probability of choosing the correct response option, and the NRM for determining the probabilities of all other response options, requiring a specification of the correct response for each item. Stewart, Drury, Wells, et al.\ used a two-dimensional NRM to explore the alignment between incorrect response options and either correct Newtonian ideas or common misconceptions on the FCI \cite{Stewart2020nrm}. Stewart, et al.\ emphasize that a benefit of using the NRM is that the correct response for each item is identified by the analysis without \textit{a priori} specification.

In this work we apply the NRM to two large data sets of student responses to the FMCE. Our goals are similar to those of Eaton, et al.\ \cite{Eaton2019irt} and Smith, et al.\ \cite{Smith2020rank} in that we want to rank the relative correctness of different response options, but we choose to approach this task without specifying the correct response, which allows us to get a better sense of how close the most highly ranked incorrect response is to being correct. We seek to answer the following research questions.
\begin{enumerate}
    \item To what degree do rankings of incorrect response options by NRM analyses (dis)agree with previously published rankings identified using the 2PL-NRM nested-logit model?
    \item How consistent are the results of NRM analyses across independent data sets?
    \item In what ways do rankings identified from pretest data differ from those identified from posttest data?
\end{enumerate}

Our initial hypothesis is that the rankings will largely be consistent across all data sets, but that differences may occur in items with  response options that are rarely chosen by students: for these responses, differences in the response patterns of a handful of students across multiple items could result in substantial shifts in the calculation of the rankings. Such differences could reveal the limitations of trying to gain information from quantitative analyses of rarely chosen responses.

The overarching goal of this project is to value the good ideas that students express when choosing response options that do not necessarily correspond with the correct Newtonian response. Analyzing data using the IRT NRM allows us to relate specific response options with students' overall knowledge and understanding of Newtonian physics. 

\section{Applying the Nominal Response Model}
Bock's NRM defines the probability of selecting response option $k$ to a particular item as,
\begin{equation}
    P(k|\theta) = \frac{e^{(a_k\theta + d_k)}}{\sum_{i=1}^{N} e^{(a_i\theta + d_i)}},\label{eqnrm}
\end{equation}
where $\theta$ is the person parameter (i.e., latent trait, or overall understanding of Newtonian concepts of force and motion). The parameters $a_k$ and $d_k$ are estimated for each response option, and the summation is taken over all responses to the item (from 5-9 options for items on the FMCE). The $a_k$ parameter is a measure of how much each response option aligns with the latent trait $\theta$, and is thus used as a metric for ranking the responses \cite{Bock2007,Smith2020rank,Eaton2019irt}. The $d_k$ parameter is related to the likelihood that students will choose each response option, independent of $\theta$; we do not use the $d_k$ values for our rankings.

For this work we use two large independent data sets. The first data set (DS1) consists of 7,288 matched (pre/post) student responses to the FMCE: 6,336 response sets were obtained from the PhysPort DataExplorer database \cite{physportde}, and 952 response sets were collected at four different colleges and universities from across the US. The second data set (DS2) consists of 6,912 response sets collected via the Learning About STEM Student Outcomes (LASSO) website \cite{lasso}. Unlike DS1, many of the response sets in DS2 include only the pre-instruction or only the post-instruction data 
\footnote{Data included in DS1 were collected before the creation of the LASSO website, so we can be sure that there is no overlap in the data sets.}. We do not consider students' matched pre/post responses in this analysis, so we include all response sets whether or not they include both pre- and post-instruction data.

We separated the data sets into four subsets: DS1pre, DS1post, DS2pre, and DS2post. For each subset we removed any response sets that included more than two blank responses. A general guideline for applying IRT analyses is that data sets should include at least 10 times as many respondents as parameters to be estimated \cite{deAyala2008}. For the FMCE, there are 361 response options across 47 items, with two parameters being estimated for each option ($a_k$ and $d_k$), meaning that a minimum of 7,220 response sets is required for each analysis. None of the four subsets contains enough data to be analyzed individually. To attempt to answer research questions 1 and 2, we combined the pre-instruction and post-instruction data from DS1 (DS1full, $N=12,388$, identical to the data set from Ref.\ \cite{Smith2020rank}) and the pre/post data from DS2 (DS2full, $N=9,875$) to create two independent data sets. To attempt to answer research question 3, we created a data set consisting of all pre-instruction responses from both DS1 and DS2 (AllPre, $N=11,494$) and a data set consisting of all post-instruction responses (AllPost, $N=10,769$). We also analyzed a data set consisting of all pre/post responses from both data sets (Combined, $N = 22,263$).

All IRT analyses were performed using the mirt package in the R computing environment \cite{mirt,r}. To get a sense of the uncertainty in the parameter values, we used the mirt function's option to generate random values for the initial parameter estimates (GenRandomPars = TRUE), and we repeated the analysis of each data set 10,000 times. We compared the resulting distribution of values for each parameter to determine the relative ordering of $a_k$ values for each item. 

As expected, the $a_k$ parameter values calculated by the mirt function for each correct response were higher than the parameters corresponding to incorrect responses for each item. When plotting the distributions of the $a_k$ values for each item, we found that at least 90\% of the analyses yielded parameters that were mostly consistent, forming a narrow distribution about a central maximum, and the rest of the analyses yielded values that were shifted higher or lower than this main group (forming a secondary distribution with the same order). 
%
To compare the values of the parameters, we identified the results contained within the main consistent group (the ``big peak'') for each item, ranked the responses based on the location of the central maximum of this distribution, and calculated the Hedges' $g$ effect size for each pair of parameter distributions. We consider responses to have approximately equal $a_k$ values if the effect size is less than 0.5, and to have meaningfully different $a_k$ values if the effect size is at least 1.3; we consider effect sizes in the range $0.5 \leq |g| < 1.3$ to represent inconclusive results.

\begin{table}
    \begin{ruledtabular}
    \begin{tabular}{crccccccccccccccc}
    Item & Data Set & \multicolumn{5}{l}{Ranking}&\\
    \hline
    17&2PL-NRM&E&$>$&F&$=$&\textbf{B}&$>$&D&$\geq$&G&$=$&C&$\geq$&H&$\geq$&A\\
    &DS1full&E&$>$&F&$>$&\textbf{B}&$>$&D&$>$&G&$=$&C&$\geq$&H&$>$&A\\
    &DS2full&E&$>$&\textbf{B}&$>$&H&$>$&A&$=$&D&$>$&G&$>$&C&$>$&F\\
    &AllPre&E&$>$&\textbf{B}&$>$&A&$>$&D&$>$&H&$>$&G&$>$&C&$>$&F\\
    &AllPost&E&$>$&\textbf{B}&$>$&H&$>$&D&$>$&A&$>$&G&$>$&C&$>$&F\\
    &Combined&E&$>$&\textbf{B}&$>$&A&$>$&D&$>$&H&$>$&G&$>$&C&$>$&F\\
    \hline
    19&2PL-NRM&B&$>$&A&$\geq$&E&$>$&\textbf{D}&$=$&F&$>$&C&$=$&G&$>$&H\\
    &DS1full&B&$>$&A&$>$&E&$>$&\textbf{D}&$=$&F&$>$&C&$\geq$&G&$>$&H\\
    &DS2full&B&$>$&A&$>$&\textbf{D}&$>$&C&$\geq$&G&$>$&H&$>$&F&$>$&E\\
    &AllPre&B&$>$&A&$>$&\textbf{D}&$>$&C&$>$&G&$>$&H&$>$&F&$>$&E\\
    &AllPost&B&$>$&A&$>$&\textbf{D}&$>$&C&$>$&G&$>$&H&$>$&F&$>$&E\\
    &Combined&B&$>$&A&$>$&\textbf{D}&$>$&C&$>$&G&$>$&H&$>$&F&$>$&E\\
    \hline
    22&2PL-NRM&A&$>$&C&$\geq$&B&$>$&D&$>$&G&$>$&F&$=$&\textbf{E}    &&\\
    &DS1full&A&$>$&C&$>$&B&$>$&D&$>$&G&$>$&\textbf{E}&$=$&F&&\\
    &DS2full&A&$>$&\textbf{E}&$>$&C&$>$&B&$>$&G&$>$&F&$>$&D&&\\
    &AllPre&A&$>$&\textbf{E}&$>$&C&$>$&G&$>$&B&$>$&F&$>$&D&&\\
    &AllPost&A&$>$&\textbf{E}&$>$&C&$>$&B&$\geq$&G&$>$&F&$>$&D&&\\
    &Combined&A&$>$&\textbf{E}&$>$&C&$>$&G&$>$&B&$>$&F&$>$&D&&\\
    \hline
    34&2PL-NRM&E&$>$&F&$=$&A&$=$&D&$=$&C&$=$&\textbf{B}&&&&\\
    &DS1full&E&$>$&F&$=$&A&$>$&D&$\geq$&C&$>$&\textbf{B}&&&&\\
    &DS2full&E&$>$&\textbf{B}&$>$&F&$>$&A&$>$&D&$>$&C&&&&\\
    &AllPre&E&$>$&\textbf{B}&$>$&F&$>$&A&$>$&D&$>$&C&&&&\\
    &AllPost&E&$>$&\textbf{B}&$>$&A&$>$&F&$>$&D&$>$&C&&&&\\
    &Combined&E&$>$&\textbf{B}&$>$&F&$>$&A&$>$&D&$>$&C&&&&\\    \end{tabular}
    \caption{Rankings identified by six different IRT analyses for four sample items. The correct response for each item is ranked highest, and the most common incorrect response is shown in bold. The 2PL-NRM rankings are copied from Ref.\ \cite{Smith2020rank}.}
    \label{tab:samplerank}
    \end{ruledtabular}
\end{table}

\begin{figure*}[tb]
    \centering
    \includegraphics[width = 0.4\textwidth]{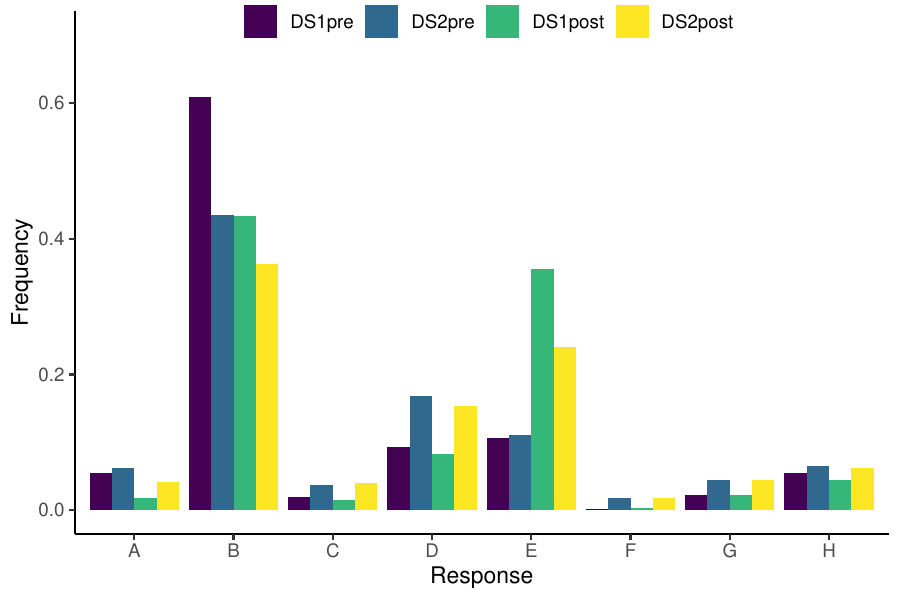}
    \hspace{0.1\textwidth}
    \includegraphics[width = 0.4\textwidth]{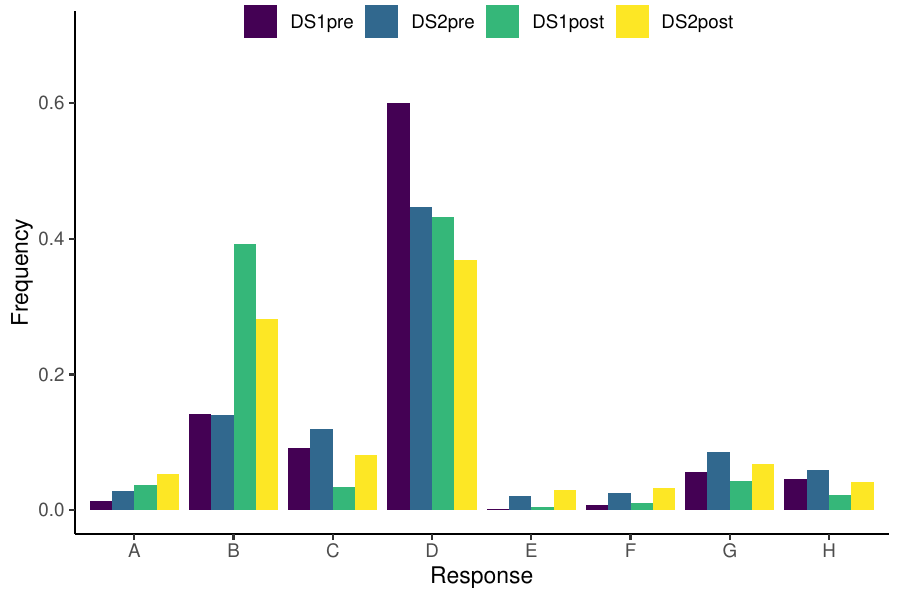}
    
    \hspace{0.21\textwidth} (a) Item 17 \hspace{0.42\textwidth} (b) Item 19 \hfill~
    
    \vspace{2mm}
    \includegraphics[width = 0.4\textwidth]{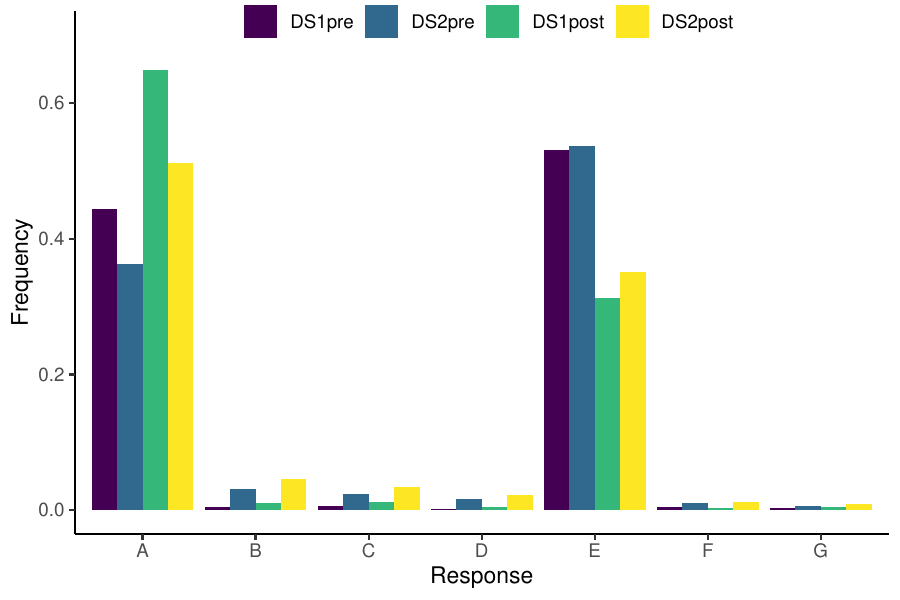}
    \hspace{0.1\textwidth}
    \includegraphics[width = 0.4\textwidth]{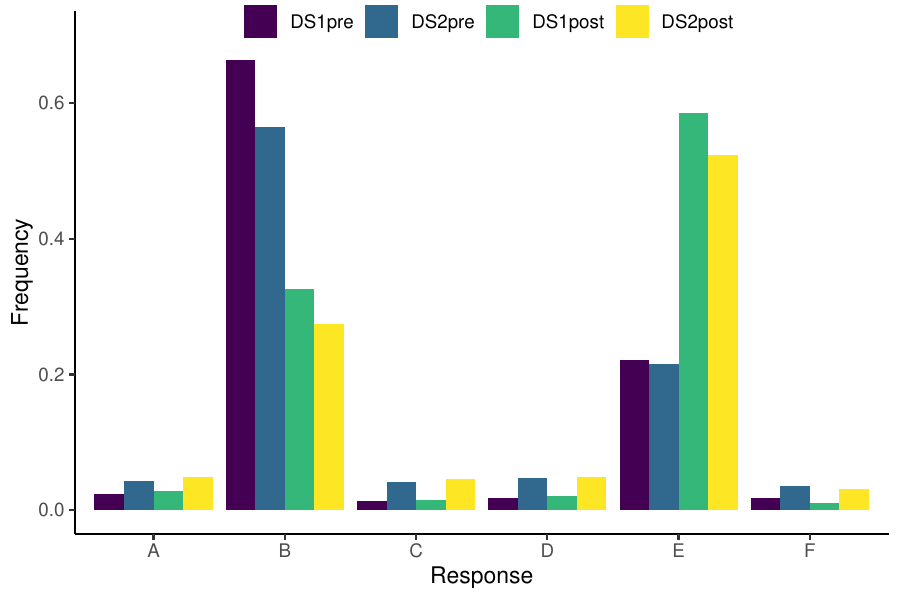}
    
    \hspace{0.21\textwidth} (c) Item 22 \hspace{0.42\textwidth} (d) Item 34 \hfill~
    \caption{Response frequencies for all four subsets of our data: DS1pre, DS2pre, DS1post, and DS2post.}
    \label{fig:responses}
\end{figure*}

    
    
    
    
    
    
    

We identified the ``big peak'' of each parameter distribution for each item in two ways. For the DS1full data set, we plotted the parameter distribution for all response options for each item and visually identified the range of values that corresponded to the central 90--95\% of the distribution for the correct response. We then filtered the analysis results to only include the results for which the $a_k$ value for the correct response was within that range (approximately 91--98\% of the 10,000 runs) and calculated Hedges' $g$ for pairs of responses. Additionally, we wrote an R script to identify the $a_k$ value that maximizes the density function for each response, define a range of values within one standard deviation of the maximum density for the correct response, select the subset of the data within the defined range, rank the responses based on the maximum density values, and calculate Hedges' $g$ for pairs of responses that are adjacent to each other in the ranking. 
The rankings identified by both of these methods agreed for all responses to all items. The pairwise Hedges' $g$ values were similar (but not identical) for all comparisons, with meaningful differences in only a handful of cases \footnote{We consider differences in $g$ to be ``meaningful'' only if they result in a different categorization of the comparison: equal, inconclusive, or different.}. We chose to use the algorithmic method for analyzing the remaining data sets (DS2full, AllPre, and AllPost) because it  reduces subjectivity in defining the range of values within the ``big peak,'' and it is much faster when we use a loop function in R. 

\section{Results}
For all items and data sets the correct response option is unambiguously determined to have the largest $a_k$ value, providing support for ranking responses by their parameter values; however, the rankings of the incorrect answers differed between the data sets for most items. We attempted to identify consistent rankings across data sets by removing less commonly chosen responses for each item. This method yielded rankings that were not very useful: rankings for 22 of the items contained only two consistently-ranked responses (the correct and one incorrect), and rankings for another 18 items had only three consistently-ranked responses.


Table \ref{tab:samplerank} shows the rankings of FMCE responses for six analyses of four sample items: the previously published 2PL-NRM rankings, as well as the rankings identified by analyzing each of our data sets using the NRM. One of the most salient results in Table \ref{tab:samplerank} is that the NRM rankings for DS1full are very similar to the previously published 2PL-NRM rankings; and the DS2full rankings are very similar to the rankings identified by analyzing the AllPre, AllPost, and Combined data sets; but these two groups are markedly different from each other. Because of these similarities, we focus our discussion on the DS1full and DS2full rankings. The four items in Table \ref{tab:samplerank} and Fig.\ \ref{fig:responses} were chosen as exemplars that are representative of trends that we see across all of the items.

For all but two (of 47) items, the most commonly chosen incorrect response is ranked higher in the DS2full rankings than it is in the DS1full rankings. In Table \ref{tab:samplerank} the most common incorrect response is shown in bold. This difference is quite stark for seven of the items (including items 22 and 34) in that the most commonly chosen incorrect response is the highest-ranked incorrect response for DS2full, but the lowest (or tied for lowest) for DS1full. Another common trend --- seen in items 17, 19, and twelve other items --- is that the response that is ranked lowest for DS2full is ranked higher than the most commonly chosen response for DS1full. Item 19 stands out in Table \ref{tab:samplerank} because the highest-ranked incorrect response is consistent in all six analyses; an additional five items had a consistent highest-ranked incorrect response, and in all cases it was not the most commonly chosen response. This consistency provides support for the argument that some items may have incorrect responses that are ``better'' than the most common response.

The fact that the AllPre, AllPost, and Combined rankings are more similar to DS2full than DS1full is particularly interesting because DS1 is a larger data set, comprising 53\% of AllPre, 59\% of AllPost, and 56\% of the Combined data set. We do not yet have a complete explanation for this phenomenon, but we believe there may be clues in the differences in overall response rates within our data sets. Figure \ref{fig:responses} shows the response rates for our four exemplar items for each of the four data subsets: DS1pre, DS1post, DS2pre, and DS2post. A common trend across all items is that students in the DS1 data set are more likely to choose the correct and most-common incorrect response than students in DS2 (e.g., responses E and B, respectively, for item 34); conversely, students in DS2 are more likely to choose less-common response options than students in DS1 (e.g., responses A, C, D, and F for item 34). 

Another trend that can be observed in Fig.\ \ref{fig:responses}(b) is that students in both DS1 and DS2 are more likely to choose the highest-ranked incorrect response to item 19 (A) after instruction than before instruction. We see similar trends for the other items with consistent highest-ranked incorrect responses. This suggests that learning may be occurring, even though these students are not choosing the correct response.

The greater likelihood of DS2 students to select less-common incorrect responses (both before and after instruction, and for all items on the FMCE) could potentially explain why the AllPre, AllPost, and Combined rankings are more similar to DS2full. Parameter values are estimated such that they represent the level of understanding of the students who choose each response; therefore, the level of understanding of an individual student will have a larger effect on the estimated parameter of a response chosen by very few students than it would on the parameter of a response chosen by many students. This is supported by the rankings of item 19: Fig.\ \ref{fig:responses}(b) shows that responses E and F are very rare for DS1 (less than 1\% each), and Table \ref{tab:samplerank} shows that, if responses E and F were removed, the rankings for DS1 and DS2 would be consistent with each other. This justification for the differences in rankings suggests that the rankings from the DS2 data set may be more representative of the overall population of introductory physics students than those from DS1.


\section{Conclusions and Future Directions}
With regard to our first research question, Table \ref{tab:samplerank} shows that the rankings from the DS1full data set are very similar to the previously published 2PL-NRM rankings. The responses are always in the same order, but some of the comparisons change from ``$\geq$'' to ``$>$'' (or similar). All of these differences show less equality between responses for the DS1full ranking; this may be the result of slightly different methods for generating uncertainty in the parameter values (Ref.\ \cite{Smith2020rank} used bootstrapped random samples, and we did not). In total we consider these rankings to be generally consistent between the 2PL-NRM and the NRM analyses, demonstrating that the NRM can in fact identify the correct response.

With regard to our second research question, our hypothesis that rankings of incorrect responses to FMCE items would be mostly consistent across data sets, and that inconsistencies could be attributed to the least-commonly chosen responses is not supported by the data. Rankings for some items behave in this fashion (item 19 is consistent except for the least-popular E and F), but this is the exception rather than the rule. Rankings for most items behave more like item 34 (the ranking of less-common responses is fairly consistent, but the most-common response is drastically different) or item 17 (rankings of incorrect responses are almost reversed between DS1full and DS2full).

According to Table \ref{tab:samplerank}, the rankings from the AllPre and AllPost data sets are mostly consistent with each other and with the rankings from the Combined data set. The highest and lowest incorrect responses are always the same, and any discrepancies are localized to two or three adjacent responses, suggesting that these responses may be more similar than suggested by the calculated Hedges' $g$ values. This suggests that rankings of incorrect responses may be stable across instruction.

These results provide several avenues for future research. Comparing Table \ref{tab:samplerank} and Fig.\ \ref{fig:responses} suggests that DS2 may be more consistent with AllPre and AllPost because the less-common responses are more popular than in DS1. This could be tested by creating a sample data set from DS1 in which the number of students selecting the most-common incorrect response on all (or most) items is reduced. Determining the effect of various aspects of the data on the rankings of incorrect response is important for having confidence in the generalizability of any rankings.  
Another direction for future research would be to utilize bootstrapping methods to simulate uncertainty in the NRM parameter estimates \cite{Smith2020rank}. Based on the results from DS1, we would not expect this to change the order of any of our parameter rankings, but it may suggest that some responses are more equivalent to each other than reported here.

Traditional instruction in physics places great value on a student's ability to determine the correct response to a given problem, especially at the introductory level. This is exacerbated by the use of multiple-choice tests where partial credit is rarely available. This focus on correctness has the potential to lead to a mentality among students that they either ``get'' physics, or they don't: a mentality that can be particularly detrimental for students who identify as members of populations traditionally underrepresented in physics. 
Ranking all response options to each item on the FMCE may lead to a data-driven model for assigning partial-credit for each response. Of course, this presupposes that a consistent ranking could be achieved that is representative of the introductory physics student population. 
The specific methods for assigning partial-credit values based on the rankings identified by the IRT NRM is beyond the scope of this paper, but we expect that future work will investigate these in greater detail.

In principle, this method for ranking incorrect responses could be applied to any multiple-choice test as long as a sufficiently large data set exists to be able to determine the initial rankings (about 20 students for every response option: e.g., 500 students for a 5-item test with five response options for each item). Combining these methods with previous uses of multidimensional IRT analyses would allow multiple scores for each test that show students their progress on each subtopic measured by the test. This would give students a much better picture of their own knowledge and understanding than a single number based solely on whether or not they provided a response corresponding with the desired Newtonian response. A more formative score report could help to reduce the sense of exclusivity in physics by focusing on what students understand, not only on what they don't understand.

\begin{acknowledgments}
	The National Science Foundation supported this work through grants DUE-1836470 and DUE-1832880. 
\end{acknowledgments}


\bibliography{references.bib}
\end{document}